# Investigation of robustness and numerical stability of multiple regression and PCA in modeling world development data

Chen Ye GAN[1,*]

[1]Department of Mathematics, University of California, Los Angeles, CA 90095
[*]felixgan@ucla.edu

**Abstract**—Popular methods for modeling data both labelled and unlabeled, multiple regression and PCA has been used in research for a vast number of datasets. In this investigation, we attempt to push the limits of these two methods by running a fit on world development data, a set notorious for its complexity and high dimensionality. We assess the robustness and numerical stability of both methods using their matrix condition number and ability to capture variance in the dataset. The result indicates poor performance from both methods from a numerical standpoint, yet certain qualitative insights can still be captured.

## 1. Introduction

The modeling of development data has been researched extensively throughout the centuries for its pragmatic uses in policy decisions. Relationships that accurately model past data and are extrapolatable for future reference are essential to understanding the developmental inclinations of a country as well as the factors that would accelerate or stifle growth. To find those relationships, models that utilize various indicators are proposed, one of which is an indicator that reflects the innovative power of a country. Economists recognize that innovation and the economic prosperity of a country goes hand in hand. One theory suggests that innovation is the essential driver of economic progress [1]. In this investigation, in light of updated data regarding the patents counts of each country throughout the last century, we will combine data for all development indicators offered by the World Bank whilst also incorporating the newest patents per country dataset to construct an accurate model of economic growth using the methods of principal component analysis (PCA) and multiple regression, and evaluate the explanatory power of each model with modeling a complex system as economic development of each country. Following conventional models, the methods discussed below will take development indicators data as input to predict nominal GDP as its sole response variable.

## 2. Methods

The data for this study covers the years between 1900 to 2021 for 101 countries. The novel data set is sourced exclusively for this study from the patent search engine PatentHub, an organization based in China which gathers patent information from major countries and international patent organization. This particular dataset pulled from PatentHub is one sorted by country and indexed by year, where the year represents the date that an accepted patent was first filed in their respective country.

  Alongside, we use the conventional country-level aggregate economic development data from the World Bank Databank [2], collected by the Development Data Group of the World Bank, as well as data contributed by the United Nations, the Organisation for Economic Co-Operation and Development, and the International Monetary Fund. The economic growth indicators for 23 countries between the years 1900 and 2020, inclusive, is used for comparison and contextualization with the patent dataset. To initialize the models, the six input indicators used are GDP adjusted for PPP, GNI per capita (current US$), Urban population percentage, Literacy rate (% of people ages 15 and above), Mortality rate – infant (per 1,000 live births), and Adjusted Net Savings (constant 2010 US$)".

*2.1 Data Cleansing*
Before running either method, the relationship between the indicators were analyzed. Using 6 indicators, each pair of relationship, and thus $\binom{6}{2}$ relationships between the indicators, were analyzed. By generating ordinary least square (OLS) plots for all relationships, their corresponding coefficient of determination ($R^2$) values are found using the standard formula below, which is then adjusted ($\overline{R^2}$) to take into account the total number of variables.

$$R^2 = 1 - \frac{RSS}{total\ sum\ of\ squares} \tag{1}$$

*where RSS stands for sum of squares of residuals, and TSS stands for total sum of squares*

$$\overline{R^2} = 1 - (1 - R^2)\left[\frac{n-1}{n-(k+1)}\right] \tag{2}$$

*where n stands for the sample size, and k stands for the number of independent variables*

which is then used to calculate the Variance Inflation Factor (VIF) for the $i^{th}$ indicator using the equation below. All relationships were found to have a VIF of less than 5, or tolerance ($\frac{1}{VIF}$) of less than 0.2, ensuring their statistical significance [3].

$$VIF_i = \frac{1}{1 - R_i^2} \tag{3}$$

Then we will test the data for normality by visualizing country data by binning and plotting their distribution, an example of which is shown in Figure 1. This visualization allows us to manually discard certain subsets of data due to issues like bimodality or insufficient data. More rigorous analysis of the shape of the data will be assessed during the fitting stage using their respective third and fourth standardized moment (assuming normally distributed), namely skewness and kurtosis, where we center the value of kurtosis to be exactly 3, where more than 3 would be classified as leptokurtic and less than 3 be platykurtic [4], where similarly skewness will be bounded by -0.5 and 0.5, outside of which we deem the data to be violating our assumption of normally distributed.

From those data, the ones closest to normal is chosen, and a Box-Cox transformation is applied before running regressions. By applying this monotonic transformation before the fitting, we allow the errors to be more normally distributed [5], producing a more stable variance, as well as increasing the symmetry of the data around its mean, allowing for better performance on both models.

*2.2 Multiple Regression*
For multiple regression, the optimization is done using OLS with respect to the level of correlation between each indicator and nominal GDP. OLS is the default method for optimizing a multiple regression model and the value it produces informs us of the weighting of that indicators in predicting, numerically, the response variable. This value will be calculated implicitly using the Python statsmodel package during its optimization stage with the formula below [6].

$$\hat{B} = (X^T X)^{-1} X^T y \qquad (4)$$

*where $\hat{B}$ is the OLS estimator, X is the aggregate matrix regressor of variable X, T the matrix transpose, and y the vector of values of the response variable*

Using multiple regression as an exploratory model to give insight into the most explanatory indicators for GDP growth, we enforce the mandatory inclusion of patent data as a necessary indicator in order to make use of the newly collected dataset. By utilizing the well-established statsmodel libraries in Python, each fitting of multiple regression using OLS will have an associated weighting table that emerges. The table below shows a sample fitting for development data for the Republic of India, where $predictor_i$ is a unique development indicator, with low to zero intercorrelation with all other predictors within a particular fitting.

Table 1. Multiple regression fitting on economic development indicators of Republic of India

|  | coef | std err | t | P>\|t\| | α < 0.025 | α > 0.975 |
|---|---|---|---|---|---|---|
| Intercept | 8.0614 | 2.146 | 3.757 | 0.001 | 3.622 | 12.501 |
| predictor_0 | 2.083e-10 | 5.61e-10 | 0.372 | 0.714 | -9.51e-10 | 1.37e-09 |
| predictor_1 | 0.0001 | 0.000 | 1.016 | 0.320 | -0.000 | 0.000 |
| predictor_2 | -0.0773 | 0.145 | -0.532 | 0.600 | -0.378 | 0.223 |
| predictor_3 | -0.0576 | 0.062 | -0.926 | 0.364 | -0.186 | 0.071 |
| predictor_4 | -0.1678 | 0.207 | -0.810 | 0.426 | -0.596 | 0.261 |
| predictor_5 | 0.1360 | 0.098 | 1.384 | 0.180 | -0.067 | 0.339 |
| predictor_6 | 0.0003 | 0.000 | 1.844 | 0.078 | -4.27e-05 | 0.001 |

These predictors will allow future studies to gain insights into the major influencing factors when determining the nominal GDP of a particular country.

Then, the numerical stability of the model to produce reasonable insights is evaluated by using the condition number of the OLS. We define a high condition number (or ill-conditioned) as larger than $10^6$ or when the matrix becomes completely singular, in which case the matrix becomes numerical instable and thus unreliable for interpolation or extrapolation.

Assuming sufficient normality of the distribution and independence of each observation, the t-distribution is used to evaluate the significance of each predictor, results of which are shown in the last two columns in Table 1. We use an underlying distribution with $H_0 = 0$ being the null hypothesis indicating no linear relationship between the development indicator and nominal GDP. The threshold for significance is set at the classic alpha = 0.05 for a two-tailed test [7].

The robustness is then evaluated by comparing the stability of the model to the skewness and kurtosis of the underlying data from which the model was generated.

To produce a comprehensive view of the robustness rather than possibly biased by a sample size too small, we used random sampling of the indicators used for each model constructed, with varying sample sizes. After filtering out indicators that were highly correlated using their relative VIFs, the remaining indicators were put into all possible sets of 4-7 indicators, of which 1000 out of the several tens of thousands of possible combinations were chosen stochastically under a uniform distribution, and the sets that yielded the best results (highest correlation value) out of the 1000 were plotted and evaluated for robustness.

*2.3 Principal Component Analysis*

PCA is implemented using the popular scikit-learn module. The data is first routinely normalized to have a mean of 0 and variance of 1 using the preprocessing module, then it is fitted for 10 principal components, after which the entire dataset is projected onto those PCs before analysis.

To evaluate the efficacy of PCA on the given dataset, a scree plot is generated by accessing the implicitly calculated "explained variance ratio" [8] and fitting those ratios onto a bar graph scaled in relation to principal component 1 (PC1). The scree plot allows us to determine the amount of variance in the dataset that is captured by each principal component starting from the first (most captured) to the tenth (least captured). Since most PCA visualizations reduce dimensions down to two, allowing for visualization on two axes, we can reasonably define robustness as the amount of variance captured on the first two PCs.

The scatter plot of the data is then generated to visualize the clustering of the dataset, where each point of the set is colored according to that country's nominal GDP. The loading scores for each factor on the first PC is also generated to compare against the weights assigned to indicators by multiple regression.

## 3. Results and Discussions

First, we assess the scatter plots of the distribution of the data. The data shown below is the data distribution from the indicator "total patent count", where each graph is a different country, titled in accordance with ISO alpha-3 country codes. Note the evidently chaotic distributions across the board, indicating the difficulty to linearly split or model the data.

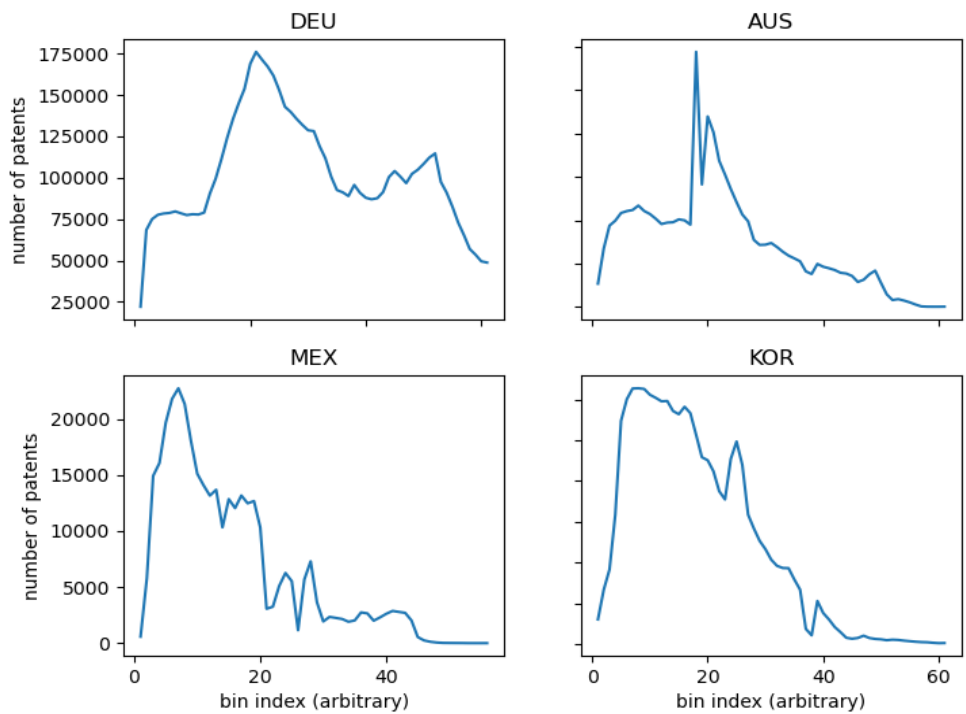

Figure 1. Total number of patents – 4 sample countries

After transformation using Box Cox power transformation, data became better and more evenly distributed. Figure 2 below shows the transformed data for KOR above.

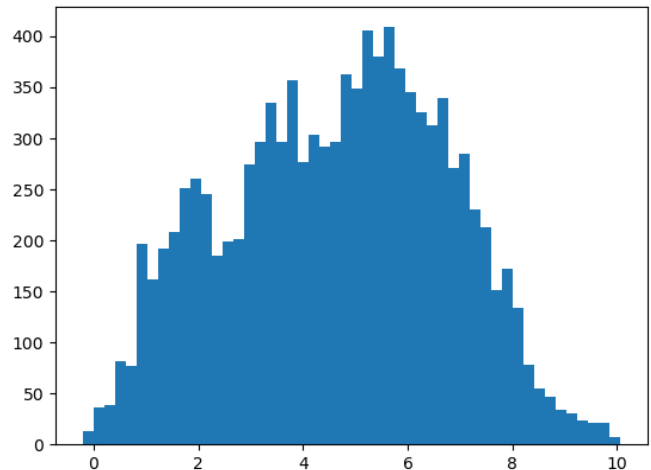

Figure 2. Binned data for KOR over 1990 – 2020 (dimensionless)

The results of multiple regression on several batch runs are listed below, along with the full generated output with various statistics. For the purpose of this investigation, we focus on three main statistics: skewness, kurtosis, condition number.

Table 2. Statistics for OLS fit on South Korea development data

| Statistic | Value |
| --- | --- |
| Omnibus | 0.645 |
| Prob(Omnibus): | 0.724 |
| **Skew:** | **0.163** |
| **Kurtosis:** | **3.128** |
| Durbin-Watson: | 1.967 |
| Jarque-Bera (JB): | 0.159 |
| Prob(JB): | 0.924 |
| **Cond. No.** | **5.79e+12** |

Table 3. Statistics for OLS fit on Top 10 GDP country aggregate development data

| Statistic | Value |
| --- | --- |
| Omnibus | 1.872 |
| Prob(Omnibus): | 0.392 |
| **Skew:** | **-0.427** |
| **Kurtosis:** | **3.231** |
| Durbin-Watson: | 1.814 |
| Jarque-Bera (JB): | 1.013 |
| Prob(JB): | 0.603 |
| **Cond. No.** | **1.42e+15** |

The statistics for the two fits listed above comes from performing OLS fit on development data of South Korea and from aggregated data of the top 10 countries with the highest nominal GDPs. Despite selecting indicators with minimal VIFs using our filtering process, and despite both skewness and kurtosis within our range of tolerance, we still achieved high condition numbers.

Customarily, when condition numbers are higher than $10^6$, the matrix is ill-conditioned, signifying that either multicollinearity has occurred between one or more indicators that were assumed to be independent of each other, or high numerical instability [9]. Since we have checked the independence of all indicators using a round of filtering with their respective VIFs, we conclude that these models constructed using multiple regression contains high numerical instabilities and thus are insufficient in explaining the fluctuations on the nominal GDP value. This could also be due to the heteroscedasticity of the data, a situation in which all the variances of the random variables are spread unequally around the response variable. This violates a basic assumption of multiple regression, and despite supposed robustness against this violation [10], results suggest otherwise when modeling global economic development data.

Moving on to PCAs, a sample generation of PCAs on all countries with data collected between the years 1990 – 2020 results in the following scree plot:

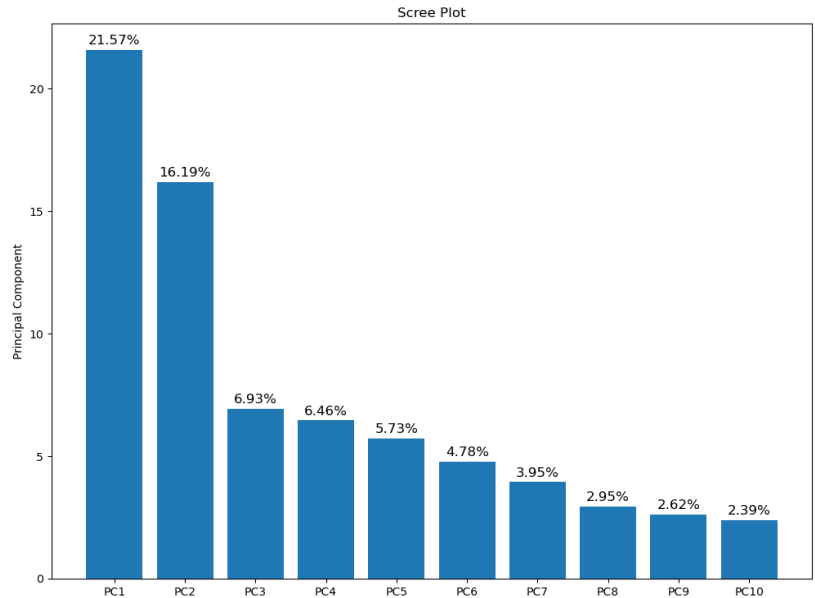

Figure 3. Scree Plot of PCA fitting on country development data

Using the criteria of limiting our dimension to two, we observe the first two PCs which add up to less than 50% of the variance explained [11]. The fat tail after the elbow of the scree plot is also indicative of more PCs needed to account for clusters that exist at those higher dimensions. This, in general, also hints at the linear inseparability of the dataset. However, even with poor robustness at only two dimensions, the scatter plot still provides insight into the dataset.

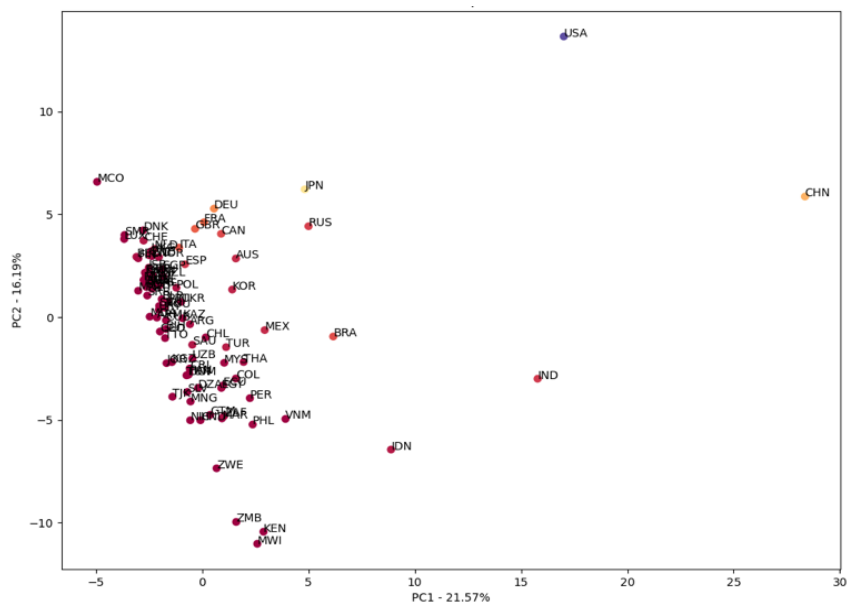

Figure 4. PCA plot of aggregate country development data

The scatterplot generated by PCA, along with coloration by nominal GDP of each country (represented by each point on the plot), shows visible clustering between countries with similar nominal GDPs, but more importantly, shows the outliers of USA, China, and India as countries with very different characteristics from other countries. This indicates that despite inability to capture all variance, some important characteristic of each country, especially major distinctions, is still captured by PCA given turbulent initial data.

Table 4. PCA loading scores on countries development data

| Predictor | Score |
| --- | --- |
| **Urban population** | **0.645** |
| **Mobile cellular subscriptions** | **0.724** |
| Agricultural nitrous oxide emissions (thousand metric tons of $CO_2$ equivalent) | 0.163 |
| Population, total | 0.195217 |
| Secondary education, general pupils | 0.195126 |

The insights generated by both graphs is compared by observing the weights of each predictor assigned by multiple regression to the loading scores by assigned by PCA. Notably in Table 1, the three predictors with highest weighting (absolute value of coefficient) for the Republic of India using OLS are *predictor$_4$*, *predictor$_5$*, *predictor$_6$*, which corresponds to *Refugee population, Arms imports, Computer/communications (% of commercial service import)*, respectively. Similarly, in Table 4, *Urban population* and *Mobile cellular subscription*, indicators comparable to *Refugee population* and *Computer/communication* in Table 1, are shown to be top predictors of development as suggested by PCA. Near identical predictors have been identified by both methods when fitted on most other subsets of this country development dataset. This high level of corroboration in the results is likely a testament to the validity of insights they individually generated. However, it could also be interpreted as an indication that both suffer from similar numerical issues or from a lack in expressive power. More research is needed to precisely determine the degree of accuracy of both methods.

**4. Conclusion**
From fitting the dataset using both multiple regression via OLS and PCA using eigenvalue optimizations, it becomes evident that given economic development data, or perhaps more generally, data with high levels of complexity and non-linear boundaries, both methods attempted shows visible limitations in reconstructing and modeling the data with high fidelity.

Multiple regression is numerically unstable despite the transformations that allowed the input data to numerically satisfy the assumptions of normality within conventionally defined boundaries. In the future, more indicators could be used to potentially allow for a more stable model. Nevertheless, the low t-statistics shows significance in the fittings, suggesting some pattern has been found. The generated weights could thus still be used in future research as reference for relative weighting between indicators.

PCA similarly suffered from a lack of robustness against economic development data. Since less than 50% of the variance is captured by the first two PCs, this suggest that the data cannot be reduced to merely two linear dimensions without losing substantial information. Possible remedies including projecting to a higher dimension under which the data becomes separable or using other non-linear dimension reduction methods that are more robust to such distributions. However, the scatter plot still managed to capture interesting characteristics of the data through its salient outliers, suggesting that some significant characteristics of a complex dataset can still be reflected at only two dimensions.